# The Evolution of tRNA-Leu Genes in Animal Mitochondrial Genomes

Paul G Higgs[1], Daniel Jameson[2], Howsun Jow[3] and Magnus Rattray[3].

1. Department of Physics, McMaster University, Hamilton, Ontario L8S 4M1, Canada.
2. School of Biological Sciences, University of Manchester, Manchester M13 9PT, UK.
3. Department of Computer Science, University of Manchester, Manchester M13 9PL, UK.

higgsp@mcmaster.ca

**Abstract**

Animal mitochondrial genomes usually have two transfer RNAs for Leucine: one, with anticodon UAG, translates the four-codon family CUN, whilst the other, with anticodon UAA, translates the two-codon family UUR. These two genes must differ at the third anticodon position, but in some species the genes differ at many additional sites, indicating that these genes have been independent for a long time. Duplication and deletion of genes in mitochondrial genomes occurs frequently during the evolution of the Metazoa. If a tRNA-Leu gene were duplicated and a substitution occurred in the anticodon, this would effectively turn one type of tRNA into the other. The original copy of the second tRNA type might then be lost by a deletion elsewhere in the genome. There are several groups of species in which the two tRNA-Leu genes occur next to one another (or very close) on the genome, which suggests that tandem duplication has occurred. Here we use RNA-specific phylogenetic methods to determine evolutionary trees for both genes. We present evidence that the process of duplication, anticodon mutation and deletion of tRNA-Leu genes has occurred at least five times during the evolution of the Metazoa - once in the common ancestor of all Protostomes, once in the common ancestor of Echinoderms and Hemichordates, once in the hermit crab, and twice independently in Molluscs.

Key Words
Mitochondrial Genomes; Phylogenetics; tRNA; Gene Duplication

**Introduction**

Almost all metazoan mitochondrial genomes possess a set of 22 transfer RNAs that is able to translate the whole of the genetic code. For most amino acids there is a single tRNA, with an anticodon able to pair with all the synonymous codons for that amino acid. This is enabled by relaxed base pairing rules at the third codon position that have been called "hyperwobble" (Kurland, 1992). In comparison, bacteria typically possess many more tRNAs: *e.g. Rickettsia prowazeckii*, the extant bacterial species thought to be most similar to the ancestral lineage from which mitochondria were derived, has 33, whilst *Mycobacterium tuberculosis* has 45 and *Escherichia coli* has 88. These figures are taken from the genomic tRNA database (Lowe & Eddy, 1997). The additional number of tRNAs in bacteria arises because there are tRNAs with different anticodons (and different codon specificities) for the same amino acid and because there are duplicate copies of genes with identical anticodons.

Leucine and Serine are the only amino acids that have six synonymous codons in the mitochondrial genetic code. Each of these requires two tRNAs, hence the total of 22 rather than 20. For Leucine, the four-codon family CUN is translated by a tRNA with anticodon UAG. We refer to the gene for this tRNA as an *L* gene in what follows. The two-codon family UUR is translated by a tRNA with anticodon UAA, and we refer to the gene for this tRNA as an *L2* gene. In a similar way, for Serine, one of the tRNAs is specific to the four-codon family and one to the two-codon family for this amino acid. In the standard genetic code, Arginine also has a four- and a two-codon family. However, in mitochondria, the two codon family is either used as stop codons or reassigned to form a four codon Serine family.

The *L* and *L2* genes on the mitochondrial genome of any one species must, by definition, differ by at least one point substitution at the third anticodon (first codon) position. In most cases, the genes differ at several other sites, which suggests that the genes have evolved independently for some time. There are constraints on tRNAs that mean that the gene sequences cannot become completely different from one another. Clearly, the secondary structure must be preserved since it is essential for the function of the molecule. It is also essential that the tRNA be recognized by the correct amino-acyl tRNA synthetase, so that it can be charged with the appropriate amino acid. It is known that certain sites in the tRNA act as identity elements that facilitate this molecular recognition process (Larkin *et al.* 2002). For Leucine tRNAs there is a single Leucyl-tRNA synthetase that functions for both the UAG and UAA anticodon tRNAs (Ribas de Pouplana & Schimmel, 2001). This enzyme must be imported into mitochondria from the cytoplasm, since mitochondrial genomes do not contain amino-acyl tRNA synthetase genes. Since the two types of tRNA-Leu are so similar in structure and function, it is likely that if a point mutation from G to A or *vice versa* occurred in the anticodon of one of the genes, the product of the mutant gene would be a functional tRNA for the alternative codon family. In other words, one type of tRNA would have evolved into the other. If this occurred, then there would now be no gene for one of the codon families and two genes for the other family. This situation would presumably be strongly deleterious and would be eliminated by selection. However, if a duplication of one of the tRNA genes occurred first, an anticodon mutation could occur in one of these duplicates, and there would still be at least one gene for each codon family. In this case the mutation would be neutral, or nearly so.

Information on gene orders in completely sequenced mitochondrial genomes has been tabulated by Boore (2000) and is also available via the web from our own relational database system OGRe (Jameson *et al.* 2003). Species with duplicate copies of genes are actually extremely rare among the metazoa. However, rearrangements of gene order are very frequent. If a region of a genome is duplicated, any genes in this region will be present twice. Mutations or deletions can then occur in either (but not both) copies of each duplicated gene, until only one copy of each gene remains. As a result of this duplication-deletion process the genes may have changed order but will have remained on the same strand. Reshuffling of gene order on the same strand is a common observation when comparing mitochondrial genomes. In the case of tRNA-Leu, the process is complicated by the fact that there are two very similar genes. Thus, our proposed mechanism for evolution of tRNA-Leu genes is: (i) duplication of one of the original two genes; (ii) anticodon mutation in one of the duplicate genes; (iii) deletion of the original copy of the non-duplicated gene. The gene deleted in step (iii) may be in a different part of the genome from the region that was duplicated. It is the aim of this paper to look for evidence that this duplication-mutation-deletion process has actually occurred.

We have recently developed a set of phylogenetic programs known as PHASE (Jow *et al.* 2002) intended specifically to study the evolution of genes for RNA molecules with conserved secondary structure, such as tRNAs and rRNAs. In this paper we use the PHASE programs to study the evolution of tRNA-Leu genes from metazoan species with complete mitochondrial genomes, paying particular attention to cases where the gene order suggests that duplications have occurred. Since we are using a single short tRNA gene to construct these phylogenies and we are looking at quite widely divergent species, we cannot expect a very good level of resolution of the trees. However, we find that in each case, the most important features of the trees come out unambiguously, and in each case there is independent information from gene order that points in the same direction as the gene sequence phylogenies. This leads us to be reasonably confident about our conclusions.

**Methods**

The species names and accession numbers for the sequences discussed in this paper are listed in Table 1. Mitochondrial gene sequences for most species in this study were downloaded from the OGRe database (Jameson *et al.* 2003; http://www.bioinf.man.ac.uk/ogre). These sequences have previously been loaded into our database from GenBank (accession numbers given in Table 1). The sequences for *Mytilus edulis, Plicopurpura columellaris* and *Littorina saxatilis* were obtained directly from GenBank. The positions of the *L* and *L2* genes in *L. saxatilis* are not annotated in the GenBank file AJ132137, however, Wilding *et al.* (1999) locate these genes as being between RNL and ND1, but in reverse order to other species, such as *P. columellaris* and *K. tunicata* (see Table 1). We used sequence alignment techniques to confirm the position of these genes in *L. saxatilis*. We also found that the genes are in reverse order and we found the start and stop positions (using the same numbering as in the GenBank file) to be: *L2* from 5708 to 5774; and *L* from 5783 to 5849.

The gene orders for the regions surrounding the *L* and *L2* genes are also given in Table 1. Full information on gene order is obtainable from OGRe (Jameson *et al.* 2003). This system allows a visual comparison of any pair of mitochondrial genomes using colour to illustrate the blocks of conserved genes between break points. As an example, Figure 1 illustrates a comparison between human and the Hemichordate, *Balanoglossus carnosus*. This example is discussed in the Deuterostomes section of the results.

Sequences of *L* and *L2* genes were aligned in a single multiple alignment using CLUSTALX (Thompson *et al.* 1997) and this was then adjusted manually with the aid of Genedoc (Nicholas & Nicholas, 1997) using inferred information about the position of the clover-leaf secondary structure. The structure is easily identifiable in most species, although details vary. The TψC loop and stem are quite variable and in some cases this stem disappears almost entirely. Alignments of subsets of species and subsets of sites were exported from the full alignment for use in phylogenetics. The alignments used contain 61 sites, consisting of all sites in the stem regions, all seven sites in the anticodon loop (including the anticodon itself), six sites from the D-loop, six sites from the multi-branched loop and an unpaired site at the 3' end. All sites in the TψC loop were excluded, since this region is extremely variable and could not be reliably aligned.

Phylogenetic trees were produced using the Markov Chain Monte Carlo (MCMC) method implemented in PHASE (Jow *et al.* 2002). A standard general reversible four state model (GR4) was used to describe sequence evolution. Variability of rate of evolution between sites was allowed for using a discrete approximation to the Gamma distribution with four rate categories (Yang, 1994). In each MCMC run, the tree topology, the branch lengths and the rate matrix parameters all vary simultaneously. Details of the proposal procedure for new tree configurations are as in Jow *et al.* (2002). A consensus tree was obtained for the set of trees generated in each MCMC run, ignoring the initial burn-in period. Repeat runs were done in order to check reproducibility of results and convergence of the MCMC sampling procedure. In order to get branch lengths on the consensus trees, the maximum likelihood tree was calculated for the consensus tree topology, keeping the topology fixed and allowing branch lengths and rate parameters to be optimized simultaneously.

In our previous papers using the PHASE package (Jow *et al.* 2002; Hudelot *et al.* 2003) we emphasized the importance of using models that account for compensatory substitutions in the paired regions of RNA helices. These models treat pairs of sites as the fundamental unit of evolution rather than single sites. The paired model we use has seven states (AU, GU, GC, UA, UG, CG, and MM). The MM state is a combination of the 10 possible mismatch base combinations that can occur. Full details of this model and of the mechanism of compensatory substitutions in RNA helices are given by Savill *et al.* (2001), Higgs (2000) and Higgs (1998). In our recent work on mammalian phyogenetics (Hudelot *et al.* 2003) we used a paired model for the helical regions and a single site model for the unpaired sites simultaneously in the likelihood calculation. However, in the present paper the amount of sequence data is limited, and trials indicated that it was not possible to reliably estimate all the parameters necessary for both models at the same time. We therefore present the results obtained using a single site GR4 model for the whole of the sequence, and ignore the issue of compensatory changes in this paper.

**Results**

*Arthropods*

The four principal Arthropod groups are the chelicerates (including the horse-shoe crab, *Limulus polyphemus*, spiders, scorpions and ticks), the myriapods (centipedes and millipedes), the insects and the crustaceans. Boore *et al.* (1998) used mitochondrial gene order to deduce the phylogeny of these four groups. They found that in several groups of Arthropods (including *L. polyphemus*, the tick *Ixodes hexagonus*, and the centipede *Lithobius forficatus*) the *L* and *L2* genes occur as neighbours between the *RNL* and *ND1* genes. This arrangement also appears in the mollusc *Katharina tunicata*, whilst in the brachiopod, *Terebratulina retusa*, and the annelid, *Lumbricus terrestris*, the gene order is very similar, except that either one or two additional tRNAs have been inserted between the *L* and *L2* genes (see Table 1, and the OGRe web site). On the other hand, in the insects and the crustaceans the two genes are in separate positions on the genome. Boore *et al.* (1998) concluded that the separation of these genes observed in crustaceans and insects is a shared derived feature that shows these two taxa are sister groups. However, if the apparent similarity of the chelicerate/myriapod order to that of the other Protostome taxa were coincidental, then the shared features of the crustacean and insect orders could be ancestral Arthropod features rather than shared derived features. We therefore wished to use phylogenetic methods to study the tRNA gene sequence evolution of the species involved in this case.

Within the Arthropods, several species stand out as having unusual mitochondrial genome rearrangements. For example, the tick *Rhipicephalus sanguineus* has a very derived gene order with respect to its relative *I. hexagonus*. The Crustaceans *Tigriopus japonicus* and *Pagurus longicarpus* are also unusual in comparison to more conservative species like *D. pulex*. In the context of tRNA-Leu genes, *P. longicarpus* merits particular attention, since the *L* and *L2* again occur as a tandem pair. In this case they are between the *COX1* and *COX2* genes, which is the standard position of the *L2* gene in most crustaceans and insects. From this, we suspected that the *L* gene had arisen by a duplication of *L2* and that the original *L* had been lost from *P. longicarpus*. The other alternative would be that the *L* gene had undergone a long-distance translocation and that it just happened to be inserted next to the *L2* gene. This seems less likely, although in fact several translocations of other tRNAs in *P. longicarpus* do seem to have occurred (Hickerson & Cunningham, 2000).

We set out to test these possibilities with molecular phylogenetics. Trees were constructed using both *L* and *L2* genes from the 13 Arthropods listed in Table 1. The tree in Figure 2 was obtained using the PHASE MCMC program. The *L* and *L2* genes are split into two well-resolved clusters with 100% posterior probability, with the exception that the *L* gene of *P. longicarpus* appears in the middle of the *L2* gene cluster, very close to the *L2* gene of the same species. This is precisely what we would expect if this were a duplicated gene.

The amount of information in a single tRNA gene is too little to resolve the phylogeny at the species level; therefore the details of the branching order within these two main groups cannot be relied on. Nevertheless, the split into the two main groups is well resolved, and given that the molecular phylogeny and the gene order information are consistent with one another, we feel confident in our interpretation with regard to *P. longicarpus*. Note that in Figure 2, the genes of the myriapod and chelicerate species (where the *L* and *L2* genes are also in tandem) separate clearly into the two groups along with the insects and the majority of the crustaceans. The typical distance between the two genes in myriapods and chelicerates is no shorter than that in insects and crustaceans and there is no suggestion of a gene duplication in myriapods and crustaceans. Thus we agree with the interpretation of Boore *et al.* (1998) that the *L,L2* pair is ancestral to Arthropods (and remains in chelicerates and myriapods), and that the *separation* of these genes is a synapomorphy for the insects and crustaceans, rather than the reverse scenario. The *L,L2* pair then reappeared as the result of a gene duplication in *P. longicarpus*.

*Molluscs*

Molluscs are another phylum where there is a mixture of species with conservative gene orders (*e.g. K. tunicata*) and derived orders (*e.g. C. gigas* and *M. edulis*). In addition to species with complete genomes, we also included the two Gastropods *P. columellaris* and *L. saxatilis*, because the partial sequences available for these species include the *L* and *L2* genes. From examination of the gene order information, two suspicious cases stand out. In *M. edulis* the *L,L2* pair precedes the *ND1* gene as in *K. tunicata* and several other protostome genomes, but it follows two other tRNAs instead of following *RNL*. There has therefore been some rearrangement going on in this area, and we must consider the possibility of gene duplications. In *L. saxatilis*, the *L* and *L2* genes appear in the usual place but in reverse order. The most likely way of switching the order of two genes is by duplication and deletion, hence this case also merits attention.

The phylogenetic tree of the Mollusc sequences obtained with PHASE is shown in Figure 3. Several Arthropod sequences were also included. The split between *L* and *L2* gene clusters is once again well resolved (94%). Three sequences appear in unusual positions. Two of these are exactly what we suspected from examining the gene order: the *L* gene of *L. saxatilis* appears in the *L2* cluster close to the *L2* gene of the same species, and the *L2* gene of *M. edulis* appears in the *L* gene cluster close to the *L* gene of the same species. These appear to be clear-cut cases of gene duplications. The third unusual case in Figure 3 is the *C. gigas L2* gene, which appears in the *L* gene cluster. However, it clusters with Arthropod *L* sequences and not with the *C. gigas L* gene. The genome of *C. gigas* has changed enormously at both gene order and sequence levels. There are very few conserved sections of gene order shared with any other species, indicating many rearrangement events peculiar to this lineage. Additionally, we have attempted phylogenetic studies of the protostome phyla using the complete set of mitochondrial proteins, and have found that *C. gigas* is very difficult to position within the tree and sometimes is not monophyletic with the other molluscs. Although there is no general consensus of the details of Mollusc phylogeny, *C. gigas* and *M. edulis* are both considered as bivalves, and they are most likely more closely related to each other than to any of the other species considered here. We believe that rapid sequence evolution and other unusual properties of the sequence are biasing our results with *C. gigas*. Given the difficulty of placing this species with a much larger protein sequence data set, it is not surprising that problems arise here when only a single short tRNA sequence is used. In fact the *L* gene of *C. gigas* appears roughly where we would expect (with other Mollusc *L* genes) although the branch length is much longer than for any other species. It is only the *L2* gene that is misplaced. Our interpretation of this is that the *L2* gene arose by duplication of the *L* gene in the bivalve lineage before the split of *M. edulis* and *C. gigas*. In the former species, the two genes remained as a tandem pair, whereas in the latter, they became separated by further genome rearrangements. The fact that the *C. gigas L2* gene appears with the Arthropod sequences would then be a phylogenetic artefact - it should really be with the Mollusc *L* genes. There are two alternative explanations that we cannot rule out: firstly, the position of the *C. gigas L2* sequence may be completely unreliable and its true position might be in the *L2* cluster (in which case no gene duplication would have happened in this species); or secondly, there might have been a gene duplication in *C. gigas* independently of the one in *M. edulis*.

*Deuterostomes*

There is one case where duplication-mutation-deletion of tRNA-Leu genes has already been suggested. Cantatore *et al.* (1987) showed that the two tRNA(Leu) genes in the sea urchin *Paracentrotus lividus* are much more similar to one another than they are in vertebrates. Furthermore, in vertebrates, the *L* gene immediately precedes the *ND5* gene, whereas in *P. lividus*, there is an additional 72 bp section at the 5' end of the *ND5* gene that has some similarity to a tRNA-Leu sequence. Hence they proposed that the *L* gene in *P. lividus* had arisen by duplication of the *L2* gene, and that the original *L* gene had become incorporated into *ND5*. Cantatore *et al.* (1987) commented that it was puzzling that the new *L* gene in *P. lividus* was widely separated from the *L2* gene, since the simplest mechanism of gaining a new tRNA would be by tandem duplication. This puzzle was resolved by the sequencing of the mitochondrial genome of the Hemichordate, *Balanoglossus carnosus* (Castresana *et al.* 1998). Hemichordates are closely related to Echinoderms and less closely to Chordates. Figure 1 shows a comparison of the genomes of *B. carnosus* and human (a representative Chordate genome). Although there are 14 break points in gene order between these species and the gene orders appear quite different at first sight, most of these differences are due to short-distance hopping of tRNA genes. If tRNAs are exluded, the only difference is the interchange of the *ND6* and *CYTB* genes. These two gene orders are substantially more similar to one another than either is to the gene orders of Echinoderms, indicating that there has been significant rearrangement in Echinoderm genomes after the split with Hemichordates. In *B. carnosus*, the *L* and *L2* genes are next to one another on the genome and also appear to be similar in sequence. The most likely explanation of all these observations is that there was a tandem duplication in the common ancestor of Echinoderms and Hemichordates, and that the two genes became separated by later rearrangement events in the Echinoderms but remained together in *B. carnosus*.

To test this, trees were constructed for the full set of Deuterostome sequences listed in Table 1 plus the sequences of the two Arthropods *Daphnia pulex* and *Locusta migratoria* as outgroups. Figure 4 shows the results from the PHASE program. The Chordate *L* genes form a well-resolved group with 100% posterior probability. A very long internal branch separates these from the large group of Deuterostome *L2* genes. The Echinoderm and Hemichordate *L* genes cluster closely with the *L2* genes and not with the Chordate *L* genes. Although the details of the branching order within these two main groups cannot be relied on, the main division in the tree is nevertheless extremely clear. Hence we conclude, in agreement with Cantatore *et al.* (1987) and Castresana *et al.* (1998), that there has been a duplication-mutation-deletion process in the common ancestor of Echinoderms and Hemichordates.

*The Ancestral Protostome*

The tree in Figure 4 has been drawn with the Arthropods as outgroup in order to illustrate the split between the two groups of deuterostome genes. However, the likelihood calculation is independent of the root position and therefore does not tell us where to place the root. Our preferred interpretation is that the root should be on the long branch separating the chordate L genes from the rest, and that the tree should be drawn as in the inset of Figure 4. There are two reasons for this. Firstly, the fact that this branch is long suggests that these groups of genes have been separate for a long time. The distance between the L and L2 genes in chordates is much longer than the equivalent distance in all the other phyla. In the echinoderms and hemichordates, this is due to a gene duplication, as described above. In the case of the protostome phyla (represented by the two arthropods in Figure 4), this can be explained by a gene duplication that occurred in the ancestral protostome.

The second reason for this interpretation comes from gene order. As described above, the arrangement *RNL,L,L2,ND1* occurs in the ancestral arthropod gene order, in *K. tunicata* (the most conserved gene order within the molluscs) and also in slightly modified form in the brachiopod, *T. retusa,* and the annelid, *L. terrestris*. Boore *et al.* (1998) give additional examples of other minor Protostome phyla with similar arrangements. We therefore believe that this arrangement is ancestral to the protostomes. However, this arrangement is not present in the ancestral deuterostome. The only deuterostome having this arrangement is *B. carnosus*, and this is as a result of a more recent duplication, as shown above. There have thus been rearrangements involving the tRNA-Leu genes along the lineage separating the ancestral deuterostome and the ancestral protostome, but we have no direct way of deducing the direction of these changes because there is unfortunately no suitable outgroup outside the bilaterians. (The Cnidarian genomes available are very divergent in gene order and do not possess tRNA genes, and so cannot be used). Thus, either a duplication occurred in the common ancestor of the protostomes, or the two genes came together by a chance translocation in the ancestor of the protostomes, or the two genes were separated by a translocation in the ancestor of the deuterostomes. However, we have already documented at least four evolutionary events where a tandem *L,L2* pair has arisen by duplication. Therefore, our interpretation is that a gene duplication occurred in the common ancestor of the protostomes.

A final point is that by placing the root on the long branch to the chordate *L* genes we are implying that the protostome *L* gene evolved by duplication of the *L2* gene, not the other way round. This same directionality is also indicated by the gene order. The arrangement *RNL,L2,ND1* occurs in many deuterostomes, and we can be fairly confident that this was in the ancestral deuterostome. Our choice of root now means that the ancestral bilaterian also contained *RNL,L2,ND1* and that duplication in the protostome lineage led to *RNL,L,L2,ND1*. There is no other positioning of the root for which all these lines of evidence can be reconciled. For example, it was suggested to us that the root should be at the node labelled 82% in Figure 4. This would mean that there was a straightforward split between *L* and *L2* genes in the deuterostomes and the duplication in the echinoderm/hemichordate lineage would not have occurred. The evidence that this duplication *did* occur is quite strong from the gene order information given above, and the remnant of the deleted gene has actually been located in *P. lividus* (Cantatore *et al.* 1987), as described above. Secondly, placing the root at the 82% node would not explain why the branch to the chordate *L* genes is so long. Thirdly, placing the root in that position would mean that the protostome *L2* gene evolved by duplication of the *L* gene. The old *L2* between *RNL* and *ND1* would then have to be deleted, and the new tandem pair would have to be coincidentally translocated back to the *same* position between *RNL* and *ND1*: this is definitely not a parsimonious explanation.

We are thus quite confident in our rooting of figure 4 on the long branch to the chordate *L* genes. From this, we conclude that the two genes present in the ancestor of all the bilaterians have survived to the present day in the chordates but in no other groups. In all the other phyla there has been at least one duplication-mutation-deletion event, and in some species there has been more than one. This explains why the two genes are more divergent from one another in chordates than in any other phylum.

**Discussion**

Figures 5 and 6 summarize our proposed scenario for the evolution of tRNA-Leu genes. Figure 5 shows a parallel gene-tree for the two genes and illustrates the position of the duplication- mutation-deletion events. Evidence for at least five such cases is given above, with the interpretation that there is a single event in the common ancestor of the two bivalves. If the apparent duplication in *C. gigas* were independent, this would be a sixth case. In four cases, the *L* gene has arisen from a duplicated *L2* gene, and in one case, the *L2* has arisen from the *L*. Figure 6, shows the species tree and uses thick lines to indicate lineages where the two genes occur as a tandem pair. In each of the proposed duplication events, the duplicated gene is created in tandem with the original gene. There are several cases where the pair is then separated by further rearrangements.

A key point in our arguments is that in species where the two genes are in tandem, the genes also tend to be more similar in sequence than would be expected. We interpreted this as evidence for gene duplications, however another explanation is possible. It is known that in genomes where recombination occurs, gene conversion can also occur, *i.e.* one sequence is replaced by a copy of the other sequence. This has been proposed as an explanation of why the different copies of genes that occur in tandem arrays (such as rRNAs in nuclear genomes) tend to be very similar in sequence to one another. It might be argued that gene conversion could occur in mitochondrial genomes when *L* and *L2* occur in tandem. However, we are not aware of any definite evidence that gene conversion occurs in mitochondria, and usually it is argued that recombination is absent or very rare in animal mitochondrial genomes, therefore we do not think this explanation is very likely. In particular, it would be difficult to explain the appearance of isolated species such as *L. saxatilis*, *M. edulis*, and *P. longicarpus*, where the genes are very similar with virtually no evolutionary time over which gene conversion might act. If gene conversion were acting, then it would have a large effect on groups like the chelicerates and myriapods where the genes have been in tandem for a long time. However, the distance between the *L* and *L2* genes in these two groups is not significantly shorter than it is for the crustaceans and insects (see Figure 2).

In a similar vein, one could ask whether horizontal gene transfer might be responsible for the unusual positioning of certain sequences in the phylogenies. However, again, we are not aware of any definite evidence that horizontal transfer occurs between animal mitochondrial genomes. Also it should be remembered that the cases we highlight here consist of genes from one family (either *L* or *L2*) that appear within the half of the tree for the *other* family. If there were horizontal transfer, then it would seem more reasonable that a gene would be replaced by a foreign gene from the *same* gene family. In this case the gene would still be in the same half of the tree. In fact we would have difficulty telling if such an event had occurred because the phylogenies are not resolved at the species level within each half of the tree. In any case, both the horizontal transfer explanation and the gene conversion explanation seem extremely unlikely in comparison to the gene duplication explanation that we give here.

In addition to the species discussed in this paper, we also examined species from several other Protostome phyla. Our results were less conclusive, because there are fewer species available in each taxon, and there are some groups, like Nematodes and Platyhelminthes, with quite derived genome sequences, making interpretation of phylogenies more difficult. We therefore do not present these results here. In the results we obtained, however, there is no evidence for any further cases of gene duplication among tRNA-Leu genes in any of the other species for which we have genomes available in the OGRe system.

The species for which we have complete mitochondrial genomes are scattered somewhat sparsely throughout the animal kingdom. As more genomes become available we expect that more cases of duplications in the tRNA-Leu genes will emerge. Since the process appears relatively common, we might ask whether the same process can occur for other types of tRNA. Serine is the other amino acid for which there are two tRNAs. However, in this case, the anticodons differ by two point mutations and it would be more difficult to convert one gene into another. We are not aware of any cases where this has happened in tRNA-Ser genes. It would be more likely that a duplicated tRNA changed into a tRNA for a different amino acid that differed by only one position in the anticodon. One case of interest is the Urochordate *Halocynthia roretzi*. This species has a variant genetic code in which AGR codons are translated as Glycine. There is an unusual tRNA-Gly gene with anticodon UCU that translates these codons in addition to the usual tRNA-Gly with anticodon UCC that translates the GGN codons. These two genes are fairly close together on the genome but are not a tandem pair. Yokobori *et al.* (1999) speculate that the additional gene could have arisen either by duplication of the original tRNA-Gly gene or by duplication of a tRNA for a different amino acid. As the database of mitochondrial genomes continues to expand, we shall be on the look out for more unusual cases of tRNA evolution.

**Acknowledgements**

P Higgs is supported by the Canada Research Chairs organization. D Jameson and M Rattray are supported by the Biotechnology and Biological Sciences Research Council and H Jow is supported by the Engineering and Physical Science Research Council.

**Figure Captions**

1. Comparison of the mitochondrial genomes of *Balanoglossus carnosus* and *Homo sapiens* produced using the OGRe system (Jameson *et al*. 2003). The colour scheme highlights the break points in gene order between the species.

2. Phylogentic tree of Arthropod tRNA-Leu genes, showing a well-resolved split between *L* and *L2* genes, with the exception that the *P. longicarpus L* gene appears in the *L2* cluster since it has arisen by a gene duplication. Posterior probabilities obtained by MCMC are given for certain clades. Most details of the arrangements at individual species level are not significant since the alignment is very short.

3. Phylogenetic tree of Mollusc tRNA-Leu genes with several Arthropods also included. The split between the *L* and *L2* genes is well-resolved, with the exception of the *L. saxatilis L* gene, which appears in the *L2* cluster, and the *L2* genes of *C. gigas* and *M. edulis*, which appear in the *L* cluster. All these cases indicate gene duplications. The latter two may be due to a single event in the common ancestor of *C. gigas* and *M. edulis*. Details of the arrangement of the species within the two clusters are not significant.

4. Phylogenetic tree of Deutersostome tRNA-Leu genes, showing that the *L* genes of Echinoderms and Hemichordates are related to the group of Deutersostome *L2* genes and not to the Chordate *L* genes. The main tree is drawn with the Arthropod sequences as outgroups. However, it is more likely that the root lies on the long internal branch between the Chordate *L* genes and the rest (as shown in the inset). This implies there have been gene duplications in the Echinoderm-Hemichordate common ancestor and also in the early Protostome lineage. Details of the arrangement of the species within the two clusters are not significant.

5. Combined Gene-tree for the *L* and *L2* genes illustrating the positions of the five documented cases of duplication, anticodon mutation, and deletion. The ancestral *L* gene (red line) survives only in Chordates, whilst he ancestral *L2* gene (black line) survives in many groups. The *L* gene arising in the Protostome ancestor (blue line) is also widespread. Green lines illustrate genes occurring only in restricted groups.

6. Species-tree for the groups that are relevant to this study. Thick lines show lineages where the *L* and *L2* genes occur as a tandem pair, and thin lines show lineages where the genes are separate. In *L. terrestris* and *T. retusa* (dashed lines) the two genes are separated by only one or two other tRNAs.

Table 1

Chordata (Vertebrata)
| | | | | |
|---|---|---|---|---|
| *Xenopus laevis* | Frog | NC 001573 | ND4,H,S2,L,ND5 | RNL,L2,ND1 |
| *Protopterus dolloi* | Lungfish | NC 001708 | as above | |
| *Cyprinus carpio* | Carp | NC 001606 | as above | |
| *Sardinonops melanosticus* | Pilchard | NC 002616 | as above | |
| *Chimaera monstrosa* | Rabbit fish | NC 003163 | as above | |
| *Squalus acanthias* | Dogfish | NC 002012 | as above | |

Chordata (Cephalochordata)
| | | | |
|---|---|---|---|
| *Branchiostoma floridae* | Lancelet | NC 000834 | as above |

Chordata (Urochordata)
| | | | | |
|---|---|---|---|---|
| *Halocynthia roretzi* | Sea squirt | NC 002177 | ND6,L,N,G,D,COX3 | ND2,H,S2,R,Q,L2,ND5 |

Hemichordata
| | | | |
|---|---|---|---|
| *Balanoglossus carnosus* | Acorn worm | NC 001887 | RNL,L,L2,ND1 |

Echinodermata
| | | | | |
|---|---|---|---|---|
| *Paracentrotus lividus* | Common urchin | NC 001572 | P,-Q,N,L,-A,W,C,-V,M,-D,Y,G,L2,ND1 | |
| *Arbacia lixula* | Black urchin | NC 001770 | as above | |
| *Strongylocentrotus purpuratus* | Purple urchin | NC 001453 | as above | |
| *Asterina pectinifera* | Starfish | NC 001627 | -ND1,-L2,-G,-Y,D,-M,V,-C,-W,A,-L,-N,Q,-P | |
| *Florometra serratissima* | Feather star | NC 001878 | P,-Q,N,L,-A,W,C,-V | -RNS,-F,-L2,-G,-RNL |

Arthropoda
| | | | | |
|---|---|---|---|---|
| *Locusta migratoria* | Locust | NC 001712 | -ND1,-L,-RNL | COX1,L2,COX2 |
| *Cochliomyia hominivorax* | Fly | NC 002660 | as above | |
| *Tribolium castaneum* | Beetle | NC 003081 | as above | |
| *Tetrodontophora bielanensis* | Springtail | NC 002735 | as above | |
| *Daphnia pulex* | Water flea | NC 000844 | as above | |
| *Penaeus monodon* | Tiger shrimp | NC 002184 | as above | |
| *Artemia franciscana* | Brine shrimp | NC 001620 | as above | |
| *Tigriopus japonicus* | Copepod | NC 003979 | COX2,L,Y,CYTB | C,A,L2,ND1 |
| *Pagurus longicarpus* | Hermit crab | NC 003058 | COX1,L,L2,COX2 | |
| *Narceus annularis* | Millipede | NC 003343 | `-ND1,-L2,-L,-RNL` | |
| *Limulus polyphemus* | Horseshoe crab | NC 003057 | as above | |
| *Ixodes hexagonus* | Hedgehog tick | NC 002010 | as above | |
| *Rhipicephalus sanguineus* | Dog tick | NC 002074 | `CYTB,S,-L,C,M` | `-ND1,-L2,-RNL` |

Mollusca

| | | | | |
|---|---|---|---|---|
| *Katharina tunicata* | Chiton | NC 001636 | `-ND1,-L2,-L,-RNL` | |
| *Loligo bleekeri* | Cephalopod (Squid) | NC 002507 | `ATP6,-H,-L,COX3` | `-ND4L,T,-L2,-G,A,D,ATP8` |
| *Pupa strigosa* | Gastropod | NC 002176 | `RNL,L,A,P,ND6` | `G,H,-Q,-L2,-ATP8` |
| *Cepaea nemoralis* | Gastropod | NC 001816 | `RNL,L,A,ND6,P` | `G,H,-Q,-L2,-ATP8` |
| *Albinaria caerulea* | Gastropod | NC 001761 | `RNL,L,P,A,ND6` | `G,H,-Q,-L2,-ATP8` |
| *Plicopurpura columellaris* | Gastropod | U29705 | `RNL,L,L2,ND1` | |
| *Littorina saxatilis* | Gastropod | AJ132137 | `RNL,L2,L,ND1` | |
| *Mytilus edulis* | Bivalve (Mussel) | M83758 | `COX2,K,M,L,L2,ND1` | |
| *Crassostrea gigas* | Bivalve (Oyster) | NC 001276 | `ND3,L,ND1` | `COX2,M,S2,L2,S,A,P,RNS` |

Other Phyla

| | | | | |
|---|---|---|---|---|
| *Terebratulina retusa* | Brachiopod | NC 000941 | `RNL,L,A,L2,ND1` | |
| *Lumbricus terrestris* | Annelid | NC 001673 | `RNL,L,A,S,L2,ND1` | |

Fig 1

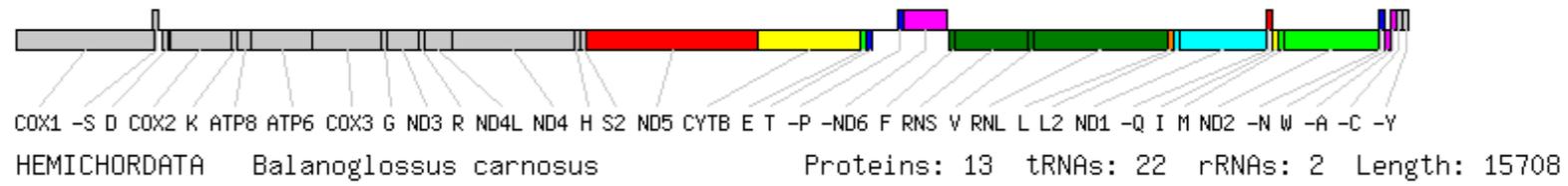

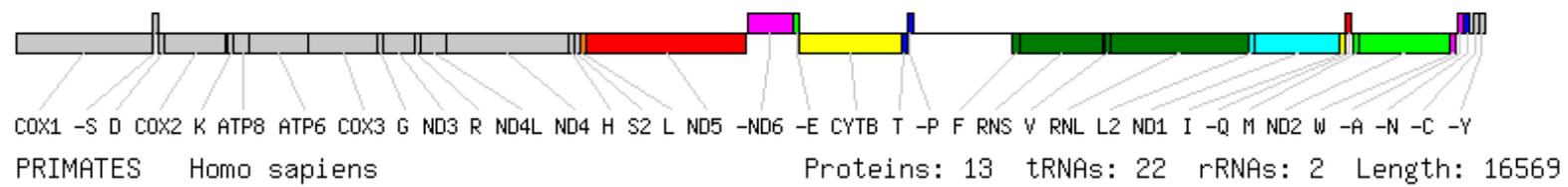

Number of Break Points: 14      tRNAs included

Fig. 2

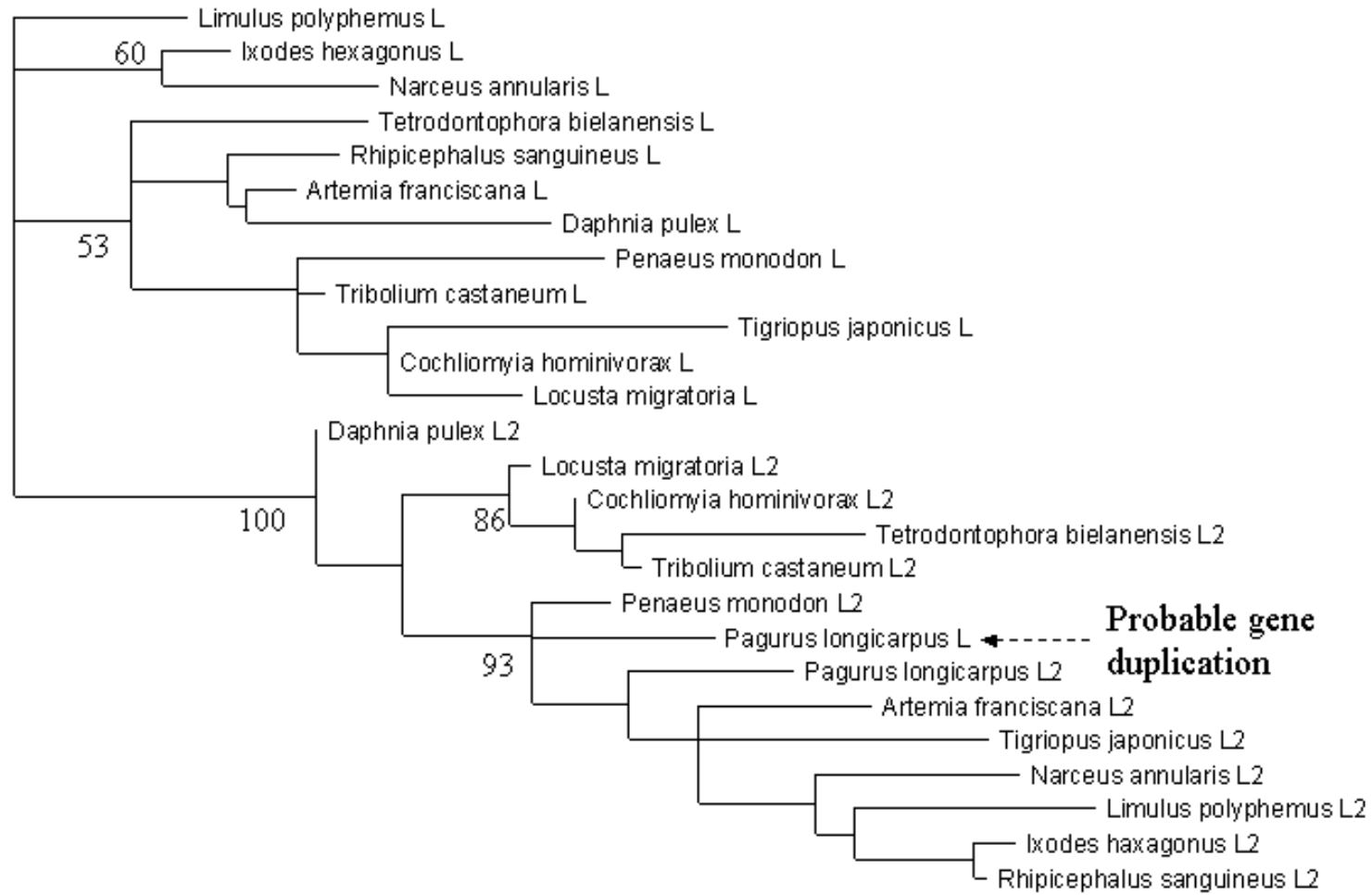

Fig 3

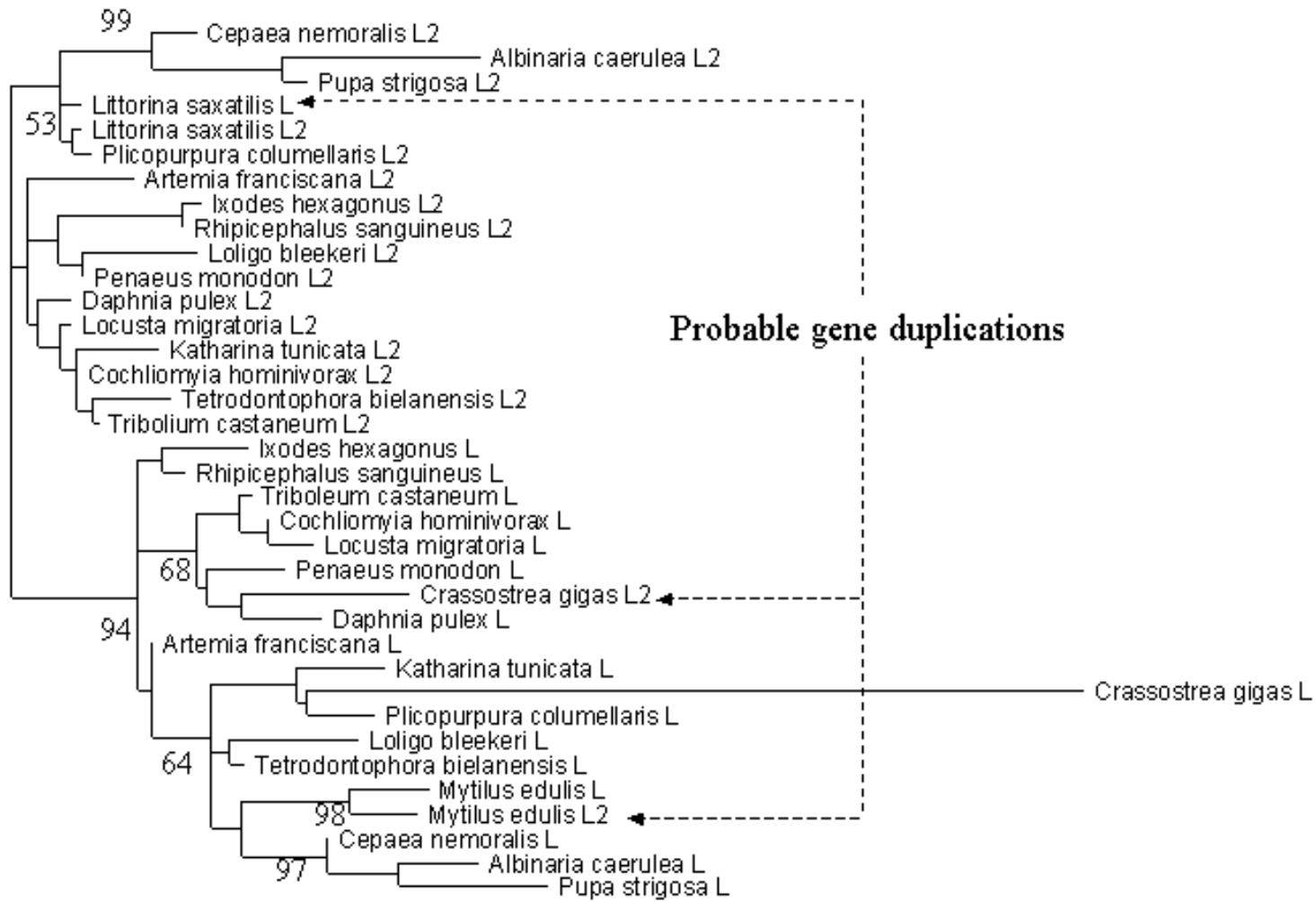

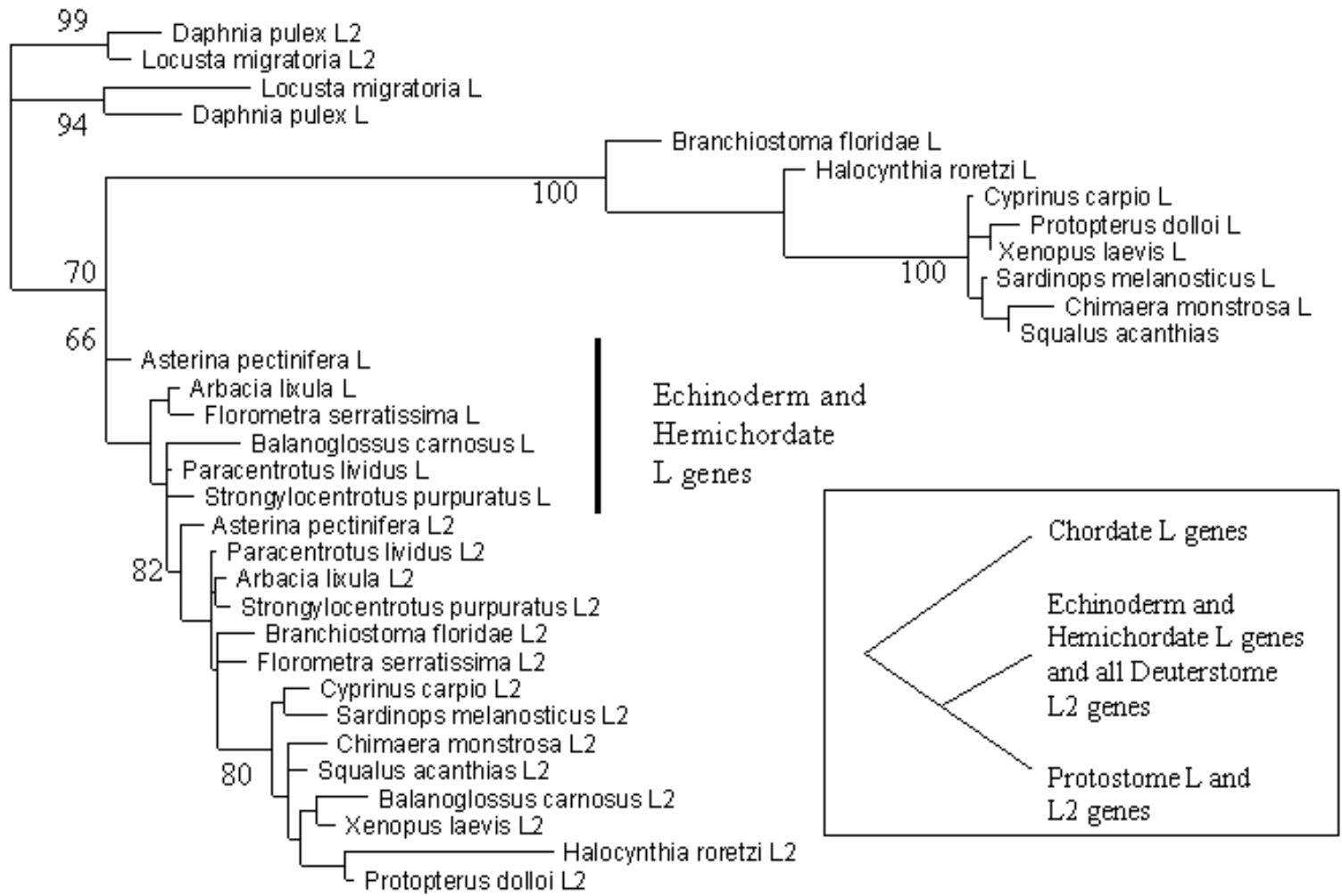

Fig 5

Fig 5. Phylogenetic tree showing L and L2 distribution across Chordates, Echinoderms and Hemichordates, Other Arthropods, *Pagurus longicarpus*, Bivalves, Other Gastropods, and *Littorina saxatilis*.

Fig 6

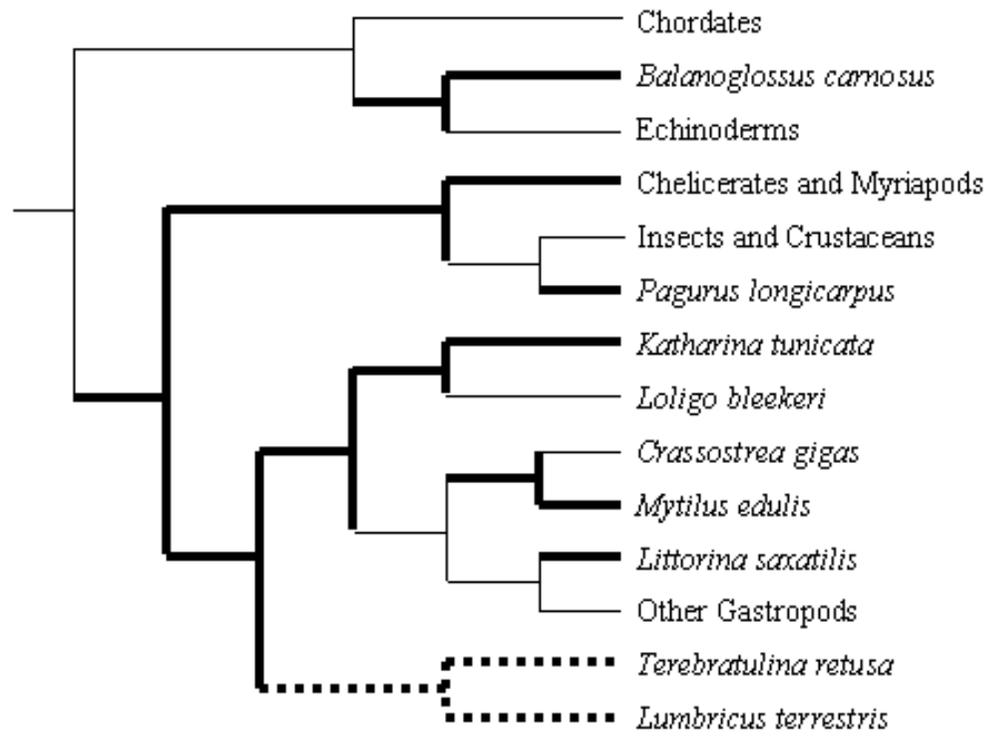